\documentclass[12pt]{article}
\usepackage{amsmath}
\usepackage{amssymb}
\usepackage[dvips]{graphicx}
\usepackage[english]{babel}

\newcommand{\be}{\begin{equation}}
\newcommand{\ee}{\end{equation}}
\newcommand{\bea}{\begin{equation}\begin{aligned}}
\newcommand{\eea}{\end{aligned}\end{equation}}

\newcommand{\eff}{\mathop{\rm eff}\nolimits}
\newcommand{\Tr}{\mathop{\rm Tr}\nolimits}
\newcommand{\lan}{\langle}
\newcommand{\ran}{\rangle}
\newcommand{\cth}{\mathop{\rm cth}\nolimits}

\begin{document}
\unitlength=1cm

\title{\bf Quark-hadron phase transition in a magnetic field}
\author{
N.~O.~Agasian\thanks{e-mail: agasian@itep.ru}~~and
S.~M.~Fedorov\thanks{e-mail: fedorov@itep.ru}
\\
{\textrm{Institute of Theoretical and Experimental Physics}} \\
{\textrm 117218 Moscow, Russia}}
\date{}

\maketitle

\begin{abstract}
Quark-hadron phase transition in QCD in the presence of magnetic field is studied. It is
shown that both the temperature of a phase transition and latent heat decrease compared
to the case of zero magnetic field. The phase diagram in the plane temperature--magnetic field is
presented. Critical point, $T_\ast=104$~MeV, $\sqrt{eH_\ast}=600$~MeV,
for which the latent heat goes to zero, is found.

\end{abstract}

\vspace{1cm}


\section{Introduction}

QCD  is essentially nonperturbative at low temperatures $T<T_c$ ($T_c$ is a quark-hadron phase transition temperature)
and is characterized by confinement and spontaneous chiral symmetry breaking. In the hadron phase, at low temperature, the
dominating contribution to the partition function of the system is given by lightest particles in the physical spectrum. In the case
of QCD these particles are $\pi$-mesons, which in the limit of massless quarks are Goldstone excitations in the chiral condensate.
It is a common method in the low temperature physics of QCD to use effective chiral theory \cite{1,2,3}, often called
chiral perturbation theory (ChPT).

One of the important questions is the phase structure of vacuum in the external magnetic field $H$.
In the recent paper \cite{Kharzeev:2007jp} it is argued that high magnetic fields $eH\sim \ 10^2 \div 10^4$~MeV$^2$
are created in heavy ion collisions. Such magnetic fields should lead to effects that can
be experimentally observed at RHIC.
Also high magnetic fields $eH\sim \Lambda_{QCD}^2$ could exist in
the early universe at the scale of strong interactions.
Such high field strength can lead to new interesting phenomena at the stage of quark-hadron phase transition.
At the same time it is interesting to study the influence of the external magnetic
field on the dynamics of strong interactions from purely  theoretical point of view.
Different nonperturbative phenomena in abelian magnetic fields were previously
studied in \cite{Shushpanov:1997sf, Agasian:1999vh,
Agasian:2000hw,  Ebert:1996tj, Agasian:1999id, Ferrer:2005vd, Ferrer:2006ie, Skalozub:1999bf,
Skalozub:2002da, Cea:2002wx, Agasian:2000cc, Kabat:2002er, Miransky:2002rp, Cohen:2007bt,
Son:2007ny, Agasian:2001ym}.

In this paper, we study the quark-hadron phase transition in QCD in abelian magnetic field. The physics of the
considered phenomenon is the following. The plasma of hot quarks and gluons at $T>T_c$ in the magnetic field
is a thermodynamic system in a paramagnetic phase. On the other hand, at low temperature, $T<T_c$, hadron matter,
which mainly consists of scalar $\pi$-mesons, is in diamagnetic phase. The paramagnetic phase is
preferable thermodynamicly, because it minimizes the free energy (maximizes pressure). Therefore the
temperature of the transition from hadron phase to the quark-gluon phase decreases as compared to
the case of zero magnetic field $H=0$. Thus, there is the analogy with the physics of condensed matter: confinement phase
corresponds to the diamagnetic gas of scalar $\pi$-mesons (we neglect the contribution of heavier hadrons), and
deconfinement phase corresponds to the paramagnetic phase of quarks and gluons.

\section{Free energy of the QCD vacuum at $T\neq0$ and $H\neq0$}

The partition function of QCD in euclidian formulation in the presence of external abelian field $A_\mu$
can be written in a form (here $T=1/\beta$ is temperature)

\be Z=\exp \left \{ -\frac{1}{4}
\int^\beta_0 dx_4\int_V d^3x F^2_{\mu\nu} \right \} \int[DB][D\bar
q][Dq] \exp \left \{ -\int^\beta_0 dx_4\int_V d^3x {\cal L} \right
\},
\label{eq_1}
\ee
where QCD Lagrangian in the background field is
\be
{\cal L}=\frac{1}{4g^2_0}
(G^a_{\mu\nu})^2+ \sum_{q=u,d} \bar q[\gamma_\mu
(\partial_\mu-iQ_q e A_\mu-i\frac{\lambda^a}{2} B^a_\mu)+m_q]q,
\label{eq_2}
\ee
here $Q_q$ is the charge matrix of quarks with flavor $q=(u,d)$, and we
omitted ghost and gauge fixing terms for simplicity. Free energy
density is given by the expression $\beta VF(T,H, m_q)=-\ln Z$.

Let us consider the hadron phase. At low temperature, $T<T_c$, ($T_c$ is the temperature of
chiral phase transition) and at weak external field, $H < \mu^2_{hadr} \sim (4\pi F_\pi)^2$, the
characteristic momentum in the vacuum loops is small, and the theory is described by the effective
low-energy Lagrangian $L_{\eff}$ \cite{2,3}.

\be
Z_{\eff}[T,H]=e^{-\beta VF_{\eff}[T,H]}=Z_0[H]\int [DU] \exp
\{ -\int^\beta_0 dx_4\int_Vd^3x L_{\eff} [U,A]\}.
\label{eq_12}
\ee

$L_{\eff}$ can be represented as a decomposition in a series of
powers of momentum (derivatives) and masses
\be
L_{\eff}=L^{(2)}+L^{(4)}+L^{(6)}+...
\label{eq_10}
\ee
The leading term in (\ref{eq_10}) is a Lagrangian of nonlinear $\sigma$-model in external
field $V_\mu$
$$
L^{(2)}=\frac{F^2_\pi}{4}\Tr(\nabla_\mu U^+\nabla_\mu U)+\Sigma
Re\Tr(\hat M U^+),
$$
\be \nabla_\mu U=\partial_\mu U-i[U,V_\mu].
\label{eq_11}
\ee
Here $U$ is unitary $SU(2)$ matrix,  $F_\pi=93 $ MeV is a pion decay constant, and the parameter $\Sigma$ has
the meaning of quark condensate, $\Sigma =|\langle \bar u u\rangle | = |\langle \bar d d\rangle |$.
External abelian magnetic field $H$, directed along $z$ axis, corresponds to $V_\mu(x)=(\tau^3/2) e A_\mu(x)$,
and vector-potential $A_\mu$ is taken in the form $A_\mu(x)=\delta_{\mu 2}Hx_1$. We will furhter
neglect the breaking of isospin symmetry of strong interactions and consider masses of light $u$- and $d$- quarks
equal, $m_u=m_d=m_q$, thus the mass matrix diagonal, $\hat M=m_q\hat I$.

In one-loop approximation it is sufficient to use the decomposition of $L_{\eff}$ up to quadratic over pion field
terms. In exponential parametrization of the matrix $U(x)= \exp \{i\tau^a\pi^a(x) /F_\pi\}$ we find that
\be
L^{(2)}=\frac12
(\partial_\mu \pi^0)+\frac12 M^2_\pi(\pi^0)^2+
(\partial_\mu\pi^++i e A_\mu\pi^+) (\partial_\mu\pi^- - i e A_\mu\pi^-)
+ M^2_\pi \pi^+\pi^-,
\label{eq_13}
\ee
where we introduced the fields of charged $\pi^\pm$ and neutral $\pi^0$ mesons
\be
\pi^\pm
=(\pi^1\pm i\pi^2)/\sqrt{2},~~ \pi^0=\pi^3.
\label{eq_14}
\ee

The QCD partition function (\ref{eq_1}) in one-loop approximation of the effective chiral
theory takes the form\footnote{The partition function $Z_{\eff}^R$ describes charged $\pi^{\pm}$
and neutral $\pi^0$ Bose gases in magnetic field}.
\be
Z_{\eff}^R[T,H]= Z^{-1}_{PT}Z_0[H]\int [D\pi^0]
[D\pi^+][D\pi^-] \exp \{-\int^\beta_0 dx_4\int_Vd^3x
 L^{(2)}[\pi, A]\}.
\label{eq_15}
\ee
Here the partition function is normalized to the case of perturbation theory at  $T=0$, $H=0$
\be
Z_{PT} = [\det(-\partial^2_\mu+M^2_\pi)]^{-3/2}.
\label{eq_16.a}
\ee
Integrating (\ref{eq_15}) over $\pi$- fields one gets
\be
Z_{\eff}^R=Z^{-1}_{PT}Z_0[H]
 [{{\rm det}_T(-\partial^2_\mu+M^2_\pi)}]^{-1/2}
 [{{\rm det}_T(-|D_\mu|^2+M^2_\pi)}]^{-1},
\label{eq_17.a}
 \ee
where $D_\mu=\partial_\mu - i e A_\mu$ is a covariant derivative, and the subscript $T$ means that the determinant
is evaluated at finite temperature $T$ using standard Matsubara rules.
Using (\ref{eq_16.a}) and regrouping multipliers in (\ref{eq_17.a}) we arrive at the following expression
for $Z_{\eff}^R$
  $$
  Z_{\eff}^R [T,H]=Z_0[H]
  \left[
   \frac{\det_T(-\partial^2_\mu+M^2_\pi)}
   {\det(-\partial^2_\mu+M^2_\pi)}\right ] ^{-1/2}
  \left[
   \frac{\det
 (-|D_\mu|^2+M^2_\pi)}
   {\det(-\partial^2_\mu+M^2_\pi)}\right ] ^{-1}
  $$
 \be
 \times \left[
   \frac{\det_T
 (-|D_\mu|^2+M^2_\pi)}
   {\det(-|D_\mu|^2+M^2_\pi)}\right ] ^{-1}.
 \label{eq_18.a}
 \ee
Free energy then takes the form \cite{Agasian:2000hw}
\be
 F_{\eff}^R(T, H)=-\frac{1}{\beta V}\ln Z_{\eff}^R
 =\frac{H^2}{2}+F_{\pi^0}(T)+F_{\pi^{\pm}}(H)
+F_{\rm dia}(T,H).
 \label{eq_16}
  \ee
Here $F_{\pi^0}$ is the free energy of massive scalar boson
\be
F_{\pi^0}(T)=T\int\frac{d^3p}{(2\pi)^3}\ln (1-\exp (-\sqrt{{\bf
p}^{2}+M^2_\pi}/T)),
\label{eq_17}
\ee
$F_{\pi^\pm}$ is Schwinger's result for the vacuum energy density of
charged scalar particles in magnetic field
\be
F_{\pi^\pm}(H)=-\frac{1}{16\pi^2}\int^\infty_0\frac{ds}{s^3}
e^{-M^2_\pi s}  \left[\frac{eHs}{\sinh(eHs)}-1\right].
\label{eq_18}
\ee

Next, it is technically not hard to generalize the case of the vacuum $H=0$, $T=0$ for charged  $\pi^\pm$-mesons
\be
F=\rm Tr
\ln(p^2_4+\omega^2_0(\bf p))
\label{eq_18a}
\ee
to the case of $ H\neq 0, T\neq 0$. Omitting the details, we will note that it corresponds to the
following substitutions
$$
p_4\to \omega_k=2\pi kT,~~~~ (k=0,\pm  1,...),
$$
\be
\omega_0=\sqrt{{\bf p}^2+M^2_\pi}
 \to \omega_n=\sqrt{p^2_z+M^2_\pi+eH(2n+1)}
\label{eq_18b}
\ee
and
$$
\rm Tr \to \frac{eHT}{2\pi} \sum^\infty_{n=0}
\sum^{+\infty}_{k=-\infty}\int^{+\infty}_{-\infty}
\frac{dp_z}{2\pi},
$$
where the degeneracy multiplicity of Landau levels $eH/2\pi$ is taken into account. Summing over Matsubara frequencies,
one obtains the following result for the diamagnetic part of free energy of charged Bose gas
$$
F_{\rm dia}(T,H)=\frac{eHT}{\pi^2}\sum^\infty_{n=0}\int^\infty_0
dk\ln (1-\exp(-\omega_n/T)),
$$
\be \omega_n=\sqrt{k^2+M^2_\pi+eH(2n+1)},
\label{eq_19}
\ee
here
$\omega_n$ are Landau levels of $\pi^\pm$-mesons in a constant magnetic field $H$.

Expanding $\ln(...)$ in the integrand of (\ref{eq_17}),
(\ref{eq_19}) in a series, one gets the following expressions:

  \be
  F_{\pi^0}=-\frac{M^2_\pi T^2}{2\pi^2}\sum^\infty_{n=1} \frac{1}{n^2}
  K_2(n\frac{M_\pi}{T})
  \label{eq_fpi0}
  \ee
and
  \be
  F_{\rm dia}=-\frac{eHT}{\pi^2}
   \sum^\infty_{n=0}
   \sqrt{M^2_\pi+eH(2n+1)}
   \sum^\infty_{k=1}\frac{1}{k}
   K_1\left ( \frac{k}{T}
   \sqrt{M^2_\pi+eH(2n+1)}\right),
    \label{eq_fdia}
   \ee
where $K_n$ is the Macdonald function.

\section{Quark-hadron phase transition in magnetic field}

In order to determine the temperature of phase transition we will write down the pressure
in two phases. At zero chemical potential pressure is equal to minus free energy.

In the confinement phase pressure of $\pi$-mesons in magnetic field
takes the form (see (\ref{eq_16}))
\be P_1(T,H)=P_{\pi^0}(T)+P_{\pi^\pm}(H) + P_{dia} (T,H),
\label{1}
\ee
where the pressure of neutral gas of $\pi^0$-mesons is $P_{\pi^0}(T)=-F_{\pi^0}(T)$.
Renormalized contribution of vacuum polarization of charged $\pi^\pm$-mesons to the pressure, which does not
depend on temperature (Schwinger polarization) is
\be
P_{\pi^\pm} (H) =\frac{1}{16\pi^2}\int^\infty_0 \frac{ds}{s^3}
e^{-M^2_\pi/s}
\left[ \frac{eHs}{\sinh(eHs)}-1+\frac16
(eHs)^2\right].
\label{3}
\ee
Diamagnetic term in the pressure, which comes from charged $\pi^\pm$ mesons is $P_{dia} (T,H)=-F_{dia} (T,H)$.

Fermion (quark) determinant at finite temperature in the magnetic field can be considered in a similar way.
Then the pressure in the quark-gluon plasma phase has the form
\be
P_{pl}(T,H) =2(N^2_c-1) \frac{\pi^2}{90}T^4 + P_q (H) +P_{para}(T,H).
\label{6}
\ee
The term proportional to $\propto T^4$ comes from the gas of hot gluons,
$P_q(H)$ is the contribution from vacuum polarization of quarks to the pressure
\be
P_q(H) =-\frac{1}{8\pi^2}\sum_{q=u,d} \int^\infty_0
\frac{ds}{s^3} e^{-m^2_qs}
\left[ |e_q|Hs \cth (|e_q|Hs) -1-\frac13
(e_qHs)^2\right]
\label{7}
\ee
and $P_{para}$ is a paramagnetic term
\be
 \begin{gathered}
P_{para} =2N_c\sum_{q=u,d} \frac{|e_q|H}{2\pi} T\sum^\infty_{n=0}
\sum_{\sigma=\pm \frac1 2}\int^{+\infty}_{-\infty}\frac{dp_z}{2\pi}
\ln (1+ e^{-\omega_q/T}),
\\
\omega_q=p^2_z+m^2_q+|e_q|H(2n+1+2\sigma).
\end{gathered}
\label{9}
\ee

An important physical property of the phase transition is the rearrangement
of the nonperturbative QCD vacuum. Due to the scale anomaly in the trace of the energy-momentum
tensor the new dimensional quantity, gluon condensate $\lan G^2\ran \equiv \lan (gG^a_{\mu\nu})^2\ran$, emerges
in QCD. Nonperturbative energy density of  the vacuum is related to $\lan G^2\ran$:
\be
\varepsilon_{vac} =-\frac{b}{128
\pi^2} \lan G^2\ran,
\label{11}
\ee
where $b=(3 N_c-2 N_f)/3,~~ N_c=3$ is the number of colors, and
$b=29/3$ for $N_f=2$.

Energy density of vacuum is the negative quantity, and the state with $\lan G^2\ran\neq 0$ turns out to be thermodynamically
advantageous. Theoretical studies \cite{Simonov:1992bc, Agasian:1993fn} and numerical computations in the
lattice approximation of QCD \cite{D'Elia:2002ck} show that at the point of phase
transition $T_c$ one part of the condensate (chromoelectric part) turns to zero, while
the chromomagnetic condensate remains almost unchanged as compared to the case $T=0$.  In the vacuum at
$T=0$  $\lan (E^a_i)^2\ran =\lan (H^a_i)^2\ran$, and therefore vacuum energy density above phase transition
appears to be less in magnitude than below phase transition, and the difference is approximately

\be
\Delta \varepsilon_v=\frac12
|\varepsilon_{vac}|=\frac{b}{256\pi^2} \lan
G^2\ran.
\label{12}
\ee

Taking this into account, the quantity $- \Delta \varepsilon_v$ should be added
to the equation of state in the plasma phase. Thus, the pressure in the quark-gluon plasma
state is given by the expression
\be
P_2(T) =P_{pl}(T,H)-\Delta\varepsilon_v.
\label{13}
\ee
Equation (\ref{13}) is  similar to the equation for the phase transition in MIT bag model,
where bag constant $B$ plays the role of  $\Delta\varepsilon_v$.

Phase transition temperature can be found from the condition of equality of pressures in both phases
 \be
 P_1(T_c,H)=P_2(T_c,H).
 \label{14}
 \ee

Let us now consider ''weak'' magnetic field, $eH\ll T_c^2$. Then the Schwinger contribution
to the pressure can be neglected. In the weak field one can use Euler-Maclaurin formula for $P_{dia}$

\be
\frac12 F(a) + \sum^\infty_{n=1} F(a+n)\approx
\int^{\infty}_{0}dx F(x) -\frac{1}{12}F'(a)
\label{41}
\ee
and
(\ref{eq_19}) can be rewritten in the form
\be
\begin{gathered}
P_{dia} =2 P_{\pi^0}(T)-
\frac{(eH)^2}{12\pi^2}h_1\left(\frac{M_\pi}{T}\right),
\\
 h_1(z)
=\int^\infty_0
\frac{dx}{\sqrt{x^2+z^2}(e^{\sqrt{x^2+z^2}}-1)}.
\end{gathered}
\label{5}
\ee
For the paramagnetic term $P_{para}$ we find in the weak field
 \be
 \begin{gathered}
P_{para}=P_{0q}(T,H=0)+\frac{N_c Q^2 (eH)^2}{6\pi^2}f_1\left(\frac{m}{T}\right)
\\
f_1(z)
=\int^\infty_0
\frac{dx}{\sqrt{x^2+z^2}(e^{\sqrt{x^2+z^2}}+1)}.
\end{gathered}
\label{10}
\ee
where $m_u=m_d=m$ and $ Q^2 =(e^2_u+ e^2_d)/e^2=(\frac49 +\frac19) =\frac59$,
and pressure $P_{0q}(T,H=0)$ is given by
\footnote{Using $\int^{\infty}_{0}
x^2 dx\ln\left( 1+e^{-x}\right)=\frac{7\pi^4}{360}$, one finds in the limit $m=0$ that
$P_{0q}(T,H=0)=4N_c\frac{7}{8}\frac{\pi^2}{90}T^4$.}

\be
\begin{gathered}
P_{0q}(T,H=0)=\frac{2N_c}{\pi^2} T^4 \int^{\infty}_{0}
x^2 dx\ln\left( 1+e^{-\omega_q/T}\right),
\\
\omega_q=\sqrt{x^2+m^2/T^2}.
\end{gathered}
\label{101}
\ee
In the absence of magnetic field and in the chiral limit one finds \cite{Simonov:1992bc}
 \be
T_c=\left(\frac{\Delta\varepsilon_v}{(\gamma-3)(\pi^2/90)}\right)^{1/4}.
\label{15}
\ee
Here $\gamma=2\cdot (N_c^2-1) + (7/8)\cdot2\cdot2\cdot N_c\cdot N_f$ is the number of
independent degrees of freedom of quarks and gluons, and  $\gamma = 37$ for $N_c=3$, $N_f=2$.
Lattice calculations give the value
$\lan G^2\ran=0.87$ $GeV^4$, and one finds from (\ref{15}) phase transition temperature  $T_c \simeq 177$ $MeV$ at $H=0$.

The influence of magnetic field can be taken into account in the first approximation by
redefining vacuum energy density
 \be
\Delta\varepsilon^H_v=\Delta\varepsilon_v-(eH)^2 V,
\label{16}
\ee
where
 \be V=\frac{1}{12\pi^2}\left[h_1\left(\frac{M_\pi}{T_c}\right) +2 N_cQ^2 f_1\left(
\frac{m}{T_c}\right)\right].
\label{17}
\ee
$V = 6.1\cdot 10^{-2}$ for $M_{\pi}=140 MeV$ and $m\approx 5 MeV$.

Considering the term $(eH)^2 V$ as a perturbation, one finds from (\ref{15}) and (\ref{16}) that
the relative shift of the deconfinement phase transition temperature in the presence of
magnetic field is
 \be
T^H_c=T_c\left( 1-\frac{V}{4\Delta \varepsilon_v}(eH)^2\right)
\label{18}
\ee
and $V/{4\Delta\varepsilon_v}\simeq 9.2 GeV^{-4}$.

Thus, we see that the presence of a magnetic field leads to a decrease of the
quark-hadron phase transition temperature, and $\Delta T/T_c \approx 10^{-2} (eH)^2/\Delta \varepsilon_v$.

\section{Results of numerical simulations}

\begin{figure}[htb]
\begin{picture}(15, 8)
\put(0.5, 0.5){\includegraphics[height=5.5cm]{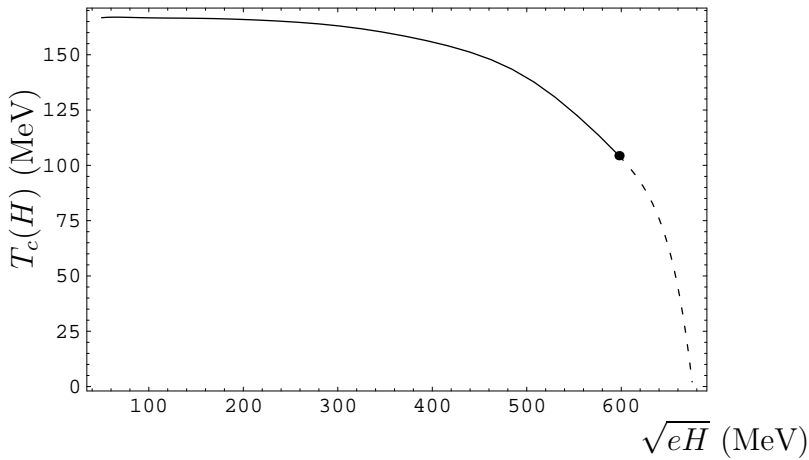}}
\put(8.5, 0.1){$\sqrt{eH}$ (MeV)}
\put(.1 , 2.5){\rotatebox{90}{$T_c(H)$ (MeV)}}
\end{picture}
\caption{Quark-hadron phase transition temperature vs external magnetic field. Critical point, $T_\ast=104$~MeV,
$\sqrt{eH_\ast}=600$~MeV, where latent heat turns to zero, is marked with the dot.
Solid line corresponds to the first order phase transition, and dashed line corresponds to the crossover.}
\label{fig_td}
\end{figure}

Equations (\ref{1}) and (\ref{13}) allow to evaluate the pressure in both phases,
and to numerically find the dependence of phase transition temperature on the magnetic field
from~(\ref{14}). Results of numerical calculations for  $N_c=3$, $N_f=2$
are presented in Fig. \ref{fig_td}. Phase transition temperature, as discussed above, decreases with
increasing external magnetic field.

Thermodynamics in each phase is defined by the pressure, and we can evaluate energy density
and latent heat in both phases. Energy density is given by
\bea
\varepsilon_1 = T \frac{dP_1}{dT} - P_1
\\
\varepsilon_2 = T \frac{dP_2}{dT} - P_2
\eea
Latent heat equals to the difference of energy densities of two phases in the point of phase transition:

\be
\Delta \varepsilon(H) = \left.\left( \varepsilon_2 - \varepsilon_1\right)\right|_{T_c(H)}
\ee

\begin{figure}[htb]
\begin{picture}(15, 8)
\put(0.5, 0.5){\includegraphics[height=5.5cm]{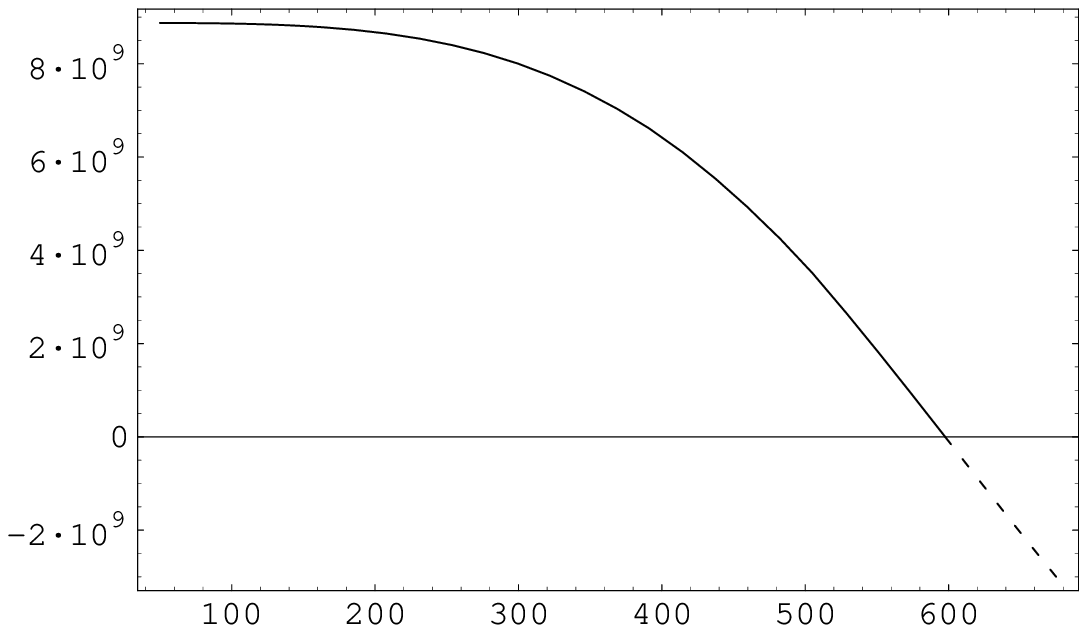}}
\put(8.5, 0.1){$\sqrt{eH}$ (MeV)}
\end{picture}
\caption{Latent heat  $\Delta \varepsilon(H)$ vs external magnetic field $\sqrt{eH}$}
\label{fig_depsilon}
\end{figure}

The dependence of $\Delta \varepsilon(H)$ is plotted in Fig. \ref{fig_depsilon}. The value of the
magnetic field $\sqrt{eH_\ast}=600$~MeV where latent heat turns to zero corresponds to the critical point,
at which first order phase transition changes to the crossover.

\section{Conclusion}

We have studied the quark-hadron phase transition in QCD in the presence of external magnetic field, and have shown,
that the temperature of the phase transition decreases in comparison to the case of zero magnetic field. Equation (\ref{14})
was solved numerically, the phase diagram in the plane temperature--magnetic field and critical point were found.

As was shown above, there are two phases in the presence of magnetic field: diamagnetic phase below $T_c$ and
paramagnetic above $T_c$. Correspondingly magnetic susceptibility, $\chi = - \partial^2 P / \partial H^2$,
changes it's sign at the critical temperature. Thus, magnetic susceptibility may be considered as
the order parameter of the model of thermal QCD in the presence of magnetic field.

It is known from lattice calculations that there is a crossover for finite temperature QCD with
physical quark masses. In the presence of magnetic field there are additional magnetic terms in
the pressure, which give different contribution to the energy density in two phases. Thus we
expect that a crossover is replaced by a first order phase transition.
Analogous phenomenon was found in~\cite{Fraga:2008qn}, where it was shown that
chiral transition changes from crossover to the weak first order transition in the linear
sigma model in a magnetic field.

The following remark should be made.
From \cite{Agasian:2000hw} it is known that the chiral phase transition temperature grows with
the magnetic field. As it follows from lattice calculations, deconfinement and chiral phase
transitions take place at the same temperature $T_c$ in case of zero magnetic field $H=0$. On the
other hand, as was shown above, quark-hadron phase transition temperature at nonzero magnetic field
is lower than in case of $H=0$. Thus, chiral and quark-hadron phase transition temperatures are
separated in the presence of magnetic field. Correspondingly, there appears a temperature
interval, where the quark-hadron phase transition is already passed, but the chiral symmetry is still
broken. This phenomenon may be important for the consideration of quark-hadron phase
transition in heavy ion collisions and in the early universe.

Authors are grateful to Yu.A.~Simonov and A.B.~Kaidalov for stimulating discussions and
useful comments.

The work  is supported
by the Federal Program of the Russian Ministry of Industry, Science, and Technology
No.40.052.1.1.1112, by the grants  of RFBR No. 06-02-17012, No. 06-02-17120
and by the grant for scientific schools NS-843.2006.2.

\end{document}